\documentclass[reprint,superscriptaddress,amsmath,amssymb,aps,prd,twocolumn,floatfix,nofootinbib]{revtex4-1}
\usepackage{amsmath, amssymb}
\usepackage{listings}
\usepackage{graphicx}
\usepackage{dcolumn}
\usepackage{bm}
\usepackage{float,color}
\usepackage{xcolor}
\usepackage[normalem]{ulem}
\usepackage{tabularx}
\usepackage{mathtools} 
\usepackage{multirow}
\usepackage{amssymb}
\usepackage{scrextend}
\usepackage{hyperref}

\begin{document}

\title{Tests of general relativity at the fourth post-Newtonian order
} 

\author{Poulami Dutta Roy}
\affiliation{Chennai Mathematical Institute, Siruseri 603103, Tamil Nadu, India}\email{poulami@cmi.ac.in}
\author{Sayantani Datta}
\affiliation{Chennai Mathematical Institute, Siruseri 603103, Tamil Nadu, India}
\affiliation{Department of Physics, University of Virginia, Charlottesville, Virginia 22904, USA}
\affiliation{Institute for Gravitation and the Cosmos, Department of Physics, Penn State University, University Park PA 16802, USA}
\author{K. G. Arun} 
\affiliation{Chennai Mathematical Institute, Siruseri 603103, Tamil Nadu, India}
\affiliation{Institute for Gravitation and the Cosmos, Department of Physics, Penn State University, University Park PA 16802, USA}

\date{\today}

\begin{abstract}
The recently computed post-Newtonian (PN) gravitational-wave phasing up to 4.5PN order accounts for several novel physical effects in compact binary dynamics such as the {\it tail of the memory, tails of tails of tails and tails of mass hexadecupole and current octupole moments}. Therefore, it is instructive to assess the ability of current-generation (2G) detectors such as LIGO/Virgo, next-generation (XG) ground-based gravitational wave detectors such as Cosmic Explorer/Einstein Telescope  and space-based detectors like LISA to test the predictions of PN theory at these orders.
Employing Fisher information matrix, we find that the projected bounds on the deviations from the logarithmic PN phasing coefficient at 4PN is ${\cal O}(10^{-2})$ and ${\cal O}(10^{-1})$ for XG and 2G detectors, respectively. Similarly, the projected bounds on other three PN coefficients that appear at 4PN and 4.5PN are ${\cal O}(10^{-1}-10^{-2})$ for XG and ${\cal O}(1)$ for 2G detectors. LISA observations of supermassive BHs could provide the tightest constraints on these four parameters ranging from ${\cal O}(10^{-4}-10^{-2})$. The variation in these bounds are studied as a function of total mass and the mass ratio of the binaries in quasi-circular orbits. These new tests are unique probes of higher order nonlinear interactions in compact binary dynamics and their consistency with the predictions of general relativity.
\end{abstract}

\maketitle
\section{Introduction}
Post-Newtonian (PN) approximation to general relativity (GR) has been very effective in modelling the compact binary dynamics during the adiabatic inspiral phase (see \cite{Blanchet:2013haa} for a comprehensive review). For non-spinning binaries in a quasi-circular orbit, the contribution to the gravitational wave phase up to 3.5PN was computed using Multipolar Post-Minkowskian formalism in Refs.~\cite{BD89,JunkS92,BS93,Blanchet:1995ez,Blanchet:1995fg,Blanchet:1996pi, Blanchet:1995ez,Blanchet:1995fg,Blanchet:1996pi,Blanchet:1996wx,Blanchet:1997jj, Blanchet:2001aw,Blanchet:2001ax,Blanchet:2004ek}. The corresponding spin-effects were computed in Refs.~\cite{KWWi93,Kidder:1995zr,Faye:2006gx,Blanchet:2006gy,Arun:2008kb,Blanchet:2011zv,Marsat:2012fn,BFH2012,M3B2013,BMB2013,Marsat2015,BoheEtal2015,Mishra:2016whh,Henry:2022dzx,Porto:2005ac,Porto:2008tb,Porto:2008jj,Maia:2017yok,Maia:2017gxn,Cho:2021mqw,Cho:2022syn}. 
Recently, the gravitational wave (GW) flux and phasing for non-spinning compact binaries was extended up to $4.5$PN, incorporating all nonlinear effects appearing till that order \cite{Foffa:2011np,Foffa:2012rn,Bini:2013zaa,Faye:2014fra,Galley:2015kus,Marchand:2016vox, Foffa:2016rgu,Porto:2017dgs,Foffa:2019rdf,Foffa:2019yfl,Blumlein:2020pog,Marchand:2020fpt,Larrouturou:2021dma,Larrouturou:2021gqo,Henry:2021cek,PhysRevD.107.104048,Blanchet:2022vsm,Trestini:2022tot,Blanchet:2023,PhysRevD.108.064041}. 

It is, therefore, pertinent to understand the importance of these newly computed terms in the context of testing GR using GWs, which forms the theme for this paper.

One of the standard methods of testing GR in the inspiral regime is the {\it parametrized tests} which are routinely performed on the GW data \cite{GW150914TGR:2016lio,GW170817TGR:2018dkp,GWTC1TGR:2019fpa,GWTC2:2020tif,GWTC3:2021sio}.
These tests make the best use of our understanding of the compact binary dynamics in GR and introduce fractional deviation parameters at different PN orders in the GW phase \cite{Arun:2004hn,Arun:2006yw,Arun:2006hn,PPE:2011ys,TIGER:2013upa,Mehta:2022pcn}. The consistency of these fractional deformation parameters with zero is assessed by measuring them from observed signals and hence are referred to as {\it null tests}. The resulting bounds from these theory-agnostic tests can be mapped to specific alternative theories of gravity as discussed, for example, in \cite{Yagi:2011xp,YYT12,YPC11,Yagi:2012ya,Yunes:2013dva,Sampson:2014qqa,Yunes:2016jcc}. The parametrized tests are currently performed up to $3.5$PN order in the inspiral phase. The newly computed $4$PN and $4.5$PN phasing corrections allow us to extend these tests and probe the novel physical effects that appear at such high PN orders and
neglect of which might result in systematic biases as shown in \cite{Owen:2023mid}.

The precision of the parametrized tests will depend on the sensitivity of the GW detector.
Proposed next-generation (XG) ground-based detectors such as Cosmic Explorer (CE) \cite{Reitze:2019iox} and Einstein Telescope (ET) \cite{Punturo:2010zz,Maggiore:2019uih} are capable of detecting compact binaries in the mass range up to a few hundreds of solar masses with a signal to noise ratio (SNR) of hundreds to thousands. Similarly, the planned Laser Interferometric Space Antenna (LISA) \cite{LISA:2017pwj} can detect mergers of supermassive black holes that have masses of the order of several millions of the solar masses, again, with SNRs of the order of thousands. Higher SNRs ensure better bounds on GR deviations. Various studies~\cite{Gupta:2020lxa,Datta:2022izc,Datta:2023muk,Hu:2022bji,Mishra:2010tp,Will:2004xi} have assessed the ability of these future detectors to carry out tests of GR. Therefore, along with the advanced LIGO (AdvLIGO) \cite{Tse:2019wcy,KAGRA:2013rdx}, advanced Virgo \cite{VIRGO:2014yos}, KAGRA \cite{KAGRA:2020agh}, GEO 600 \cite{Luck:2010rt}  and LIGO-India \cite{LIGO_India,Saleem:2021iwi}, future GW detectors can test GR with unprecedented 
precision which should be explored in the context of the new PN terms introduced in the inspiral phase.  

\raggedbottom

\subsection{Structure of the newly computed PN coefficients} 

The post-Newtonian theory is used to get an analytical expression of the inspiral GW phase in the slow-motion, weak-field regime when $v/c\ll1$ and the binary constituents are sufficiently far away from each other. Within the framework of PN theory, in order to calculate the GW phase analytically, the binding energy ($E$) and GW flux ($\mathcal{F}$) emitted by the inspiralling binaries are expressed as a series in $v/c$, the structure of which, in geometrical units, can be schematically written as
\begin{eqnarray}
    E = -\frac{1}{2} \eta v^2 \sum_{k=0}^{N} E_k v^{k}, \,\, \mathcal{F} = \frac{32}{5} \eta^2 v^{10} \sum_{k=0}^N \mathcal{F}_k v^k , \label{eq:flux}
\end{eqnarray}
where $E_k$ and $\mathcal{F}_k$ are the PN expansion coefficients that appear in the energy and flux, respectively. For non-spinning binaries, these are functions of $\eta$, the symmetric mass ratio which is related to mass ratio $q =\frac{m_1}{m_2}$ by $\eta=\frac{m_1m_2}{(m_1+m_2)^2} = \frac{q}{(q+1)^2}$ ($m_1$ and $m_2$ denote the masses of the individual components of the binary). We will follow the convention $m_1 \geq m_2$ and $G=c=1$ throughout the paper.

In the {\it adiabatic approximation}, the energy balance equation, $-dE/dt = \mathcal{F}$, in conjunction with the binding energy and flux functions introduced earlier, help us compute the phase evolution $\Phi (t)$ of the GW signal. 
One can use the stationary phase approximation (SPA) \cite{DIS00,Cutler_Flanagan} to perform the Fourier transform of the time domain gravitational wave signal and derive the phase (and amplitude) in the frequency domain for the ($\ell=2, m=2$) mode considered here with aligned spins. This, until 3.5PN, is a power series in $v$ and $\ln v$, where $v= (\pi \,\rm{M} \,f)^{1/3}$ is characteristic orbital velocity of the binary. The structure of the phase reads as
\begin{eqnarray}
    \Phi_{\rm insp}(f)&=&2\pi f t_c-\phi_c+\frac{3}{128\eta v^5}\sum_{k=0}^{7}\left[\phi_k\,v^k+\phi_{\rm k\ell}v^k\ln v\right.\nonumber\\
    &&+\left.\phi_{{\rm k}l^2} v^k\ln^2 v+\cdots\right].\label{eq:PNphase}
\end{eqnarray}
In the expression above, $t_c$ and $\phi_c$ are two kinematical parameters that denote the time of coalescence and phase of coalescence. The leading order contribution (referred to as Newtonian or 0PN) corresponds to $k=0$ and any term corresponding to $v^{k}$ will be referred to as $\frac{k}{2}$ PN, in our notation. 

Newly computed terms at 4PN and 4.5PN add to this structure. In order to highlight the structure of the new phasing terms, we re-write Eq.(\ref{eq:PNphase}) as
\begin{eqnarray}
    \Phi_{\rm insp}(f)& =& 2\pi f\,t_c-\phi_c+\frac{3}{128\eta v^5}\left[\phi_{\rm {3.5 PN}} \right.\nonumber\\
    &&+ v^8 \left(\phi_{\rm 8\ell}\ln v + \phi_{\rm 8\ell^2} \ln^2 v\right)\nonumber\\
    &&\left.+ v^9\left( \phi_{\rm 9} + \phi_{\rm 9\ell} \ln v\right)\right], \label{eq:4PNphase}
\end{eqnarray}
where $\phi_{\rm {3.5 PN}}$ denotes the 3.5PN phasing, normalized to the leading order Newtonian term, and the other terms denote the new PN coefficients at 4PN and 4.5PN orders.
The explicit expressions of the PN coefficients $\phi_{\rm 8\ell},\,\phi_{\rm 8\ell{^2}},\,\phi_{\rm 9}$ and $\phi_{\rm 9\ell}$ can be found in \cite{Blanchet:2023,PhysRevD.108.064041}. 
Until 3.5PN, that is ${\cal O}(v^7)$, the phasing in the frequency domain contains powers of $v$ and two logarithmic terms at 2.5PN and 3PN. The logarithmic term at 2.5PN is not a generation effect (such a term does not appear in the GW flux), but a consequence of the SPA. The non-logarithmic terms at 2.5PN can be reabsorbed into a redefinition of $\phi_c$. The new terms at 4PN and 4.5PN bring two new logarithmic terms at 4PN and 4.5PN as well as a $\ln^2 v$ at 4PN, apart from a non-logarithmic term at 4.5PN. There also exists a non-logarithmic term at 4PN which can be absorbed into a redefinition of $t_c$. Apart from the non-spinning terms, starting at 1.5PN, the GW phasing contains spin effects like spin-orbit and spin-spin coupling along with tail-induced spin effects. Such effects are known completely for quasi-circular orbits with non-precessing spin till 3.5PN order~\cite{KWWi93,Kidder:1995zr,Faye:2006gx,Blanchet:2006gy,Arun:2008kb,Marsat:2012fn,Mishra:2016whh,Blanchet:2011zv,BFH2012,BMB2013,M3B2013,Marsat2015,BoheEtal2015,Henry:2022dzx}. At 4PN, the next-to-next-to-leading order contribution of spin-spin interaction is also known \cite{Cho:2022syn}. We incorporate these spin effects in the inspiral phase up to 4PN order. 

\subsection{Physical effects at the new PN orders}
Each PN order in phase carries signatures of various physical effects, which become more evident when the GW flux $\mathcal{F}$ (Eq.\ref{eq:flux}) is expanded in terms of radiative multipole moments of the source~\cite{Th80} as 
\begin{eqnarray}
    \mathcal{F} = \sum_{\ell \geq 2} \big[a_\ell U_L U_L + b_\ell V_L V_L], 
\end{eqnarray}
where $U_L$ and $V_L$ denote multi-index symmetric trace free tensors that represent the mass and current radiative multipole moments of the compact binary (see Eq.~(2.1) in \cite{PhysRevD.108.064041}) with $a_\ell$ and $b_\ell$ being numerical coefficients. Each PN term in flux would contain information about corresponding multipole moments upto certain PN orders~\cite{Blanchet:1995fg,Blanchet:1996pi,Blanchet:1996wx,Blanchet:1997jj,Blanchet:2001aw,Blanchet:2001ax,Blanchet:2023,Blanchet:2013haa}. 
For example,
the computation of the flux at 4PN would require the knowledge of the mass quadrupole contribution $U_{ij}$ computed till 4PN, mass octupole $U_{ijk}$ and current quadrupole $V_{ij}$ till 3 PN, mass hexadecapole moment $U_{ijkl}$ and current octupole $V_{ijk}$ till 2 PN, moments $U_{ijklm}$, $V_{ijkl}$ at 1 PN and finally $U_{ijklmn}$, $V_{ijklm}$ at Newtonian order. The relation between radiative multipoles and source multipoles contain several nonlinear effects of GR such as tails~\cite{BD92,BS93,Blanchet:1995fg} and memory~\cite{Chr91,Th92,Arun:2004ff}. 

At $1.5$PN in the flux, the GW `tail' effect first appears, corresponding to the quadratic interaction between static ADM mass and (source-type) mass quadrupole moment~\cite{BD92,BS93,Blanchet:1995fg}. Physically, it denotes the back-scattering of the quadrupolar GW by the spacetime curvature generated by the source's ADM mass. It is a `hereditary' effect due to its dependence on the entire history of the source till the retarded time. Similarly, at $2.5$PN in the polarization, the `memory' effect appears~\cite{Chr91,Th92,Arun:2004ff}, which corresponds to quadrupole-quadrupole interaction (re-radiation of the stress-energy tensor). However, in the flux this is reduced to an instantaneous term due to the derivative operation. With increasing PN order, the complexity of the radiative moments increase as they contain higher order PN corrections to the existing effects as well as new nonlinear interactions which have been studied in detail in literature till $3.5$PN~\cite{BD92,Blanchet:1997ji,PhysRevD.101.064033,Blanchet:1997jj,Faye:2014fra,PhysRevD.107.104048,Marchand:2016vox,Blanchet:2022vsm}.

At the newly computed $4$PN order \cite{Blanchet:2023,PhysRevD.108.064041}, two novel physical effects appear for the first time, namely (i) `tail-of-memory' and (ii) `spin-quadrupole tail' both of which are hereditary effects. The tail-of-memory term denotes the scattering of re-radiated radiation by the background curvature of the source while the spin-quadrupole tail corresponds to the scattering of the radiation emitted from spin-quadrupole interaction. A quartic interaction, dubbed `tails-of-tails-of-tails', occurs at $4.5$PN order along with quartic memory interactions. Testing the agreement of such higher-order PN terms with GR provides a unique opportunity to quantify the consistency of novel physical effects occurring at these orders with the GW signal.

\subsection{Parametrized tests of GR} \label{subsec:TGR}
The elegant structure of the PN phasing formula provides the perfect testing ground to  probe the validity of GR through the parametrized tests~\cite{Arun:2006yw,Arun:2006hn,LIGOScientific:2016lio}.
These theory-agnostic tests of GR introduce normalized deviation parameters at each PN order of the inspiral phase. The coefficient at each PN order, $\phi_a$ where $a=\{\rm k, k\ell, k\ell^2\}$ and denote the non-logarithmic, logarthmic and square-logarithmic parts of the PN phase, is modified with a fractional deformation parameter $\delta \hat{\phi}_a$ ($\delta \hat{\phi}_{\rm k}$, $\delta \hat{\phi}_{\rm k\ell}$ and $\delta \hat{\phi}_{\rm k\ell^2}$) such that $\phi_a \rightarrow \phi_a^{\rm GR}(1+\delta \hat{\phi}_a)$.
By definition, $\delta\hat\phi_a=0$ denotes GR and if the posterior distribution of these parameters for a compact binary signal is consistent with zero, one would argue that the signal is statistically consistent with GR predictions. One can combine the information about these parameters from multiple events which, if GR is true, will help us place more stringent constraints than the individual events.
The state-of-the-art bounds from deviations from GR for PN orders from -1PN till 3.5PN with LIGO/Virgo detectors can be found in Fig. 6 and 7 of \cite{GWTC3:2021sio}.

\begin{figure}[h]
    \centering
    \includegraphics[width=0.48\textwidth]{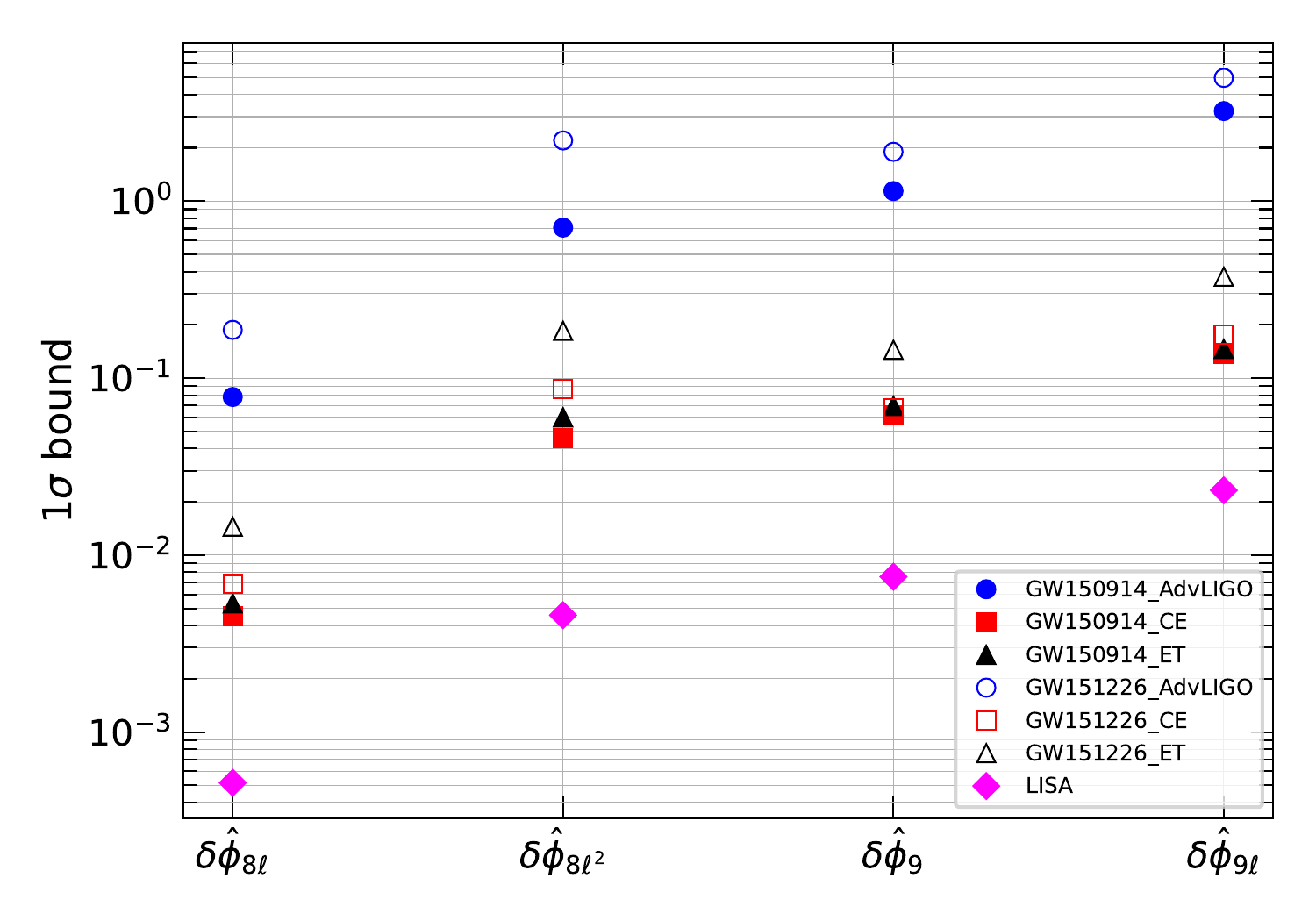}
    \caption{Projected $1\sigma$ bounds on the four new deformation parameters introduced at 4 and 4.5PN for a GW150914-like (at 440 Mpc) and GW151226-like (at 450 Mpc) system with the sensitivities of AdvLIGO, CE and ET. Among ground-based detectors, CE/ET provides a tighter constraint than AdvLIGO for a particular system. Note that the bounds for a binary computed with CE and ET are comparable. The binary for LISA has total mass $10^6$ $M_{\odot}$, $q=1.2$, aligned spins of (0.2,0.1) at 3 Gpc. The best bound on parameters are obtained from supermassive binary black holes observed in LISA. } 
    \label{fig:summary}
\end{figure}

In the spirit of the parametrized tests, we can introduce two null parameters each at 4 and 4.5PN orders. At 4PN, there will be a logarithmic ($\delta \hat{\phi}_{\rm 8\ell}$) and logarithmic-square term ($\delta \hat{\phi}_{\rm 8\ell^2}$). At 4.5PN there is a a non-logarithmic ($\delta \hat{\phi}_{\rm 9}$) and a logarthmic term ($\delta \hat{\phi}_{\rm 9\ell}$).

As the measurement of all of these parameters are accompanied by statistical uncertainties arising from the detector noise, we need to have a computationally inexpensive tool which can forecast the projected bounds on them in a reasonably reliable manner. Fisher information matrix~\cite{Cramer46,Rao45,Helstrom68,Cutler_Flanagan,Clifford_Will} provides such a semi-analytical tool which can estimate the projected bounds in the limit of sufficiently high SNR and is discussed in detail in Sec. \ref{sec:Fisher}. 

The future GW detectors, as discussed earlier, are expected to provide more stringent bound on the deviation parameters due to their enhanced sensitivity and hence higher SNR.
In this work, we employ the Fisher matrix to compute the bounds on the four new deviation parameters introduced at 4PN and 4.5PN using the noise power spectral densities (PSD) of the current (LIGO/Virgo) and XG GW detectors (Cosmic Explorer, Einstein Telescope and LISA). 

A summary of our results can be found in Fig.~\ref{fig:summary} where we provide the projected bounds on the four new deformation parameters at 4 and 4.5PN for the noise PSDs of AdvLIGO, CE, ET and LISA. For the ground based detectors, we choose  GW150914-like and GW151226-like systems as shown in Fig.(\ref{fig:summary}) while for LISA we consider a binary of mass $10^6$ $M_{\odot}$, mass ratio $1.2$, spins of magnitude $(0.2,0.1)$ at luminosity distance of 3 Gpc. The 4PN log term $\delta \hat{\phi}_{8\ell}$ is seen to be best bounded irrespective of the detector and all the deformation parameters have best constraint from supermassive binary black holes observed in LISA. Note that the bounds projected with CE and ET sensitivities are comparable and so, we consider CE as a representative of the XG detectors while computing bounds for most cases. However, it is known that ET has lower cut-off frequency smaller than CE while CE has more sensitivity in the frequency band of 10 to 200 Hz. This trade-off might influence our results when studying the entire parameter range of BBH masses. Hence, we make a comparison of the bounds from CE and ET for certain parameter values to ensure our conclusions remain consistent (see Fig.\ref{fig:Compare}).

The remainder of the paper is organized as follows. In Section \ref{sec:waveform}, we briefly discuss the waveform model used in our analysis and the deformation coefficients introduced at 4 and 4.5PN. Section \ref{sec:Fisher} explains the formalism of the Fisher information matrix used to compute the $1\sigma$ bounds on the deformation coefficients. The main result obtained in our work, i.e. the bounds on the deformation parameters, are discussed in Section \ref{sec:result} followed by the conclusion in Section \ref{sec:conclusion}. In appendix \ref{sec:appendixb}, we provide an assessment of how far the Fisher-based projections may be from the actual errors based on some representative binaries that have been detected and analysed.

\section{Waveform Model} \label{sec:waveform}
It is important to employ accurate waveform models for efficient and unbiased parameter inference. The advances in numerical relativity (NR) (see Ref.~\cite{Pretorius07Review} for a review) have made it possible to construct phenomenological waveforms that include the inspiral and merger of binary compact objects, followed by the ringdown of the remnant formed. Such waveforms are often referred to as IMR waveforms. An important subclass of waveforms called {\tt IMRPhenom}~\cite{Ajith:2007qp} was constructed to obtain a semi-analytical, computationally efficient waveform family suitable for GW searches and parameter estimation. Initially developed only for binaries with spins aligned with the orbital angular momentum vector \cite{Khan:2015jqa,Pratten:2020fqn}, they were later modified to include precession \cite{Khan:2018fmp,Pratten:2020ceb} and higher modes\cite{Garcia-Quiros:2020qpx}. 

As the real GW signals will have an inspiral, merger and ringdown, our parametrization should be on the inspiral part of an IMR waveform to avoid any biases. For our purposes, we find it sufficient to use a non-precessing phenomenological family of waveforms called {\tt IMRPhenomD}~\cite{Khan:2015jqa}. The {\tt IMRPhenomD} waveform is based on a combination of analytic post-Newtonian and effective-one-body (EOB) methods describing the inspiral regime and calibration of the merger-ringdown model to numerical relativity simulations. Hence, it is easy to construct a parametrized IMR model where any of the PN coefficients are deformed from the GR value via the parametrization discussed earlier (see Sec.\ref{subsec:TGR}). As the detected population of compact binaries to date is dominantly non-precessing~\cite{GWTC3:2021sio}, the projected bounds should still be representative of what may be achieved. A future work that assesses these bounds within the framework of Bayesian inference should employ more up-to-date waveforms with higher modes and precession effects such as {\tt IMRPhenomXPHM} \cite{Pratten:2020ceb} .

Schematically the frequency domain {\tt IMRPhenomD} waveform can be written as
\begin{align}
    \Tilde{h}(f) = \mathcal{A}(f) \, e^{i \Phi (f)}  
\end{align}
where $\mathcal{A}(f)$ and $\Phi(f)$ are the amplitude and phase of the waveform. The amplitude in the inspiral part agrees with the standard PN phase given in Eq.(\ref{eq:PNphase}) up to 3.5PN order. We modify the inspiral segment of the {\tt IMRPhenomD} waveform to incorporate the 4PN and 4.5 PN phasing terms as described in Eq.(\ref{eq:4PNphase}). We also introduce the four new deformation parameters \{$\delta \hat{\phi}_{\rm 8\ell}, \delta \hat{\phi}_{\rm 8\ell^2},\delta \hat{\phi}_{\rm 9}, \delta \hat{\phi}_{\rm 9\ell}$\} in the inspiral phase of the waveform. We have removed the non-logarithmic terms occurring at $2.5$PN and $4$PN as they can be absorbed in the re-definition of $\phi_c$ and $t_c$ respectively.

Ideally, the deformation parameters occurring at all PN orders should be measured simultaneously since any putative GR violation can occur at any PN order which priorly is not known. However, due to the strong correlation among the deformation parameters themselves and also with the GR parameters, such multi-parameter tests are uninformative, leading to poor estimation of the deviation parameters. Hence, one resorts the obvious alternative of performing single-parameter tests where each deformation parameter is estimated at a time, along with other GR parameters of the binary. This has become a norm in tests of GR using gravitational waves. (See for instance \cite{Arun:2006yw,Pai:2012mv,Datta:2020vcj,Gupta:2020lxa,Saleem:2021nsb, Datta:2022izc,Datta:2023muk} where multi-parameter tests are discussed in detail).

In this work, we will also restrict to the standard practice of performing single-parameter tests where one of these deformation parameters are estimated along with all the GR parameters ${\mathbf {\theta}_{\rm GR}}=\{\rm{ln}\, \rm{d_L},t_c,\phi_c,\rm{ln}\, M_c, \eta, \chi_1,\chi_2\}$ where $\chi_{1,2}$ denote the dimensionless spin parameters of the binary components, $\rm{d_L}$ is the luminosity distance of the binary and $\rm{M_c}$ is the chirp-mass related to the total mass $\rm{M}$ by $\rm{M_c} = \rm{M} \, \eta^{3/5}$. Therefore, the $7+1$ dimensional parameter space consists of 7 GR parameters and one deformation parameter. 

Given the designed noise PSD of a GW detector, an estimate of the $1\sigma$ error bars associated with measuring these parameters can be obtained via Fisher information matrix~\cite{Cutler_Flanagan,Clifford_Will}. 
Since we are interested to study the bounds on the PN deformation parameters and their correlations with the intrinsic parameters, we do not consider the effects of sky localisation and orientations. The averaging over the source location and orientation results in a pre-factor of $2/5$ multiplied to the amplitude of the waveform \cite{FinnCh93,Robson:2018ifk} for the case of AdvLIGO, CE and ET. To include the triangular shape of ET, a factor of $\sqrt{3}/2$ is multiplied to the waveform amplitude. On the other hand, the noise PSD of LISA already takes into account the $60^{\rm o}$ angle between the detector arms and the sky location and polarization averaging factors\cite{Babak:2017tow,Datta:2023muk}. Hence, while computing the bounds for LISA, only a factor of $\sqrt{4/5}$ is multiplied to the amplitude of the {\tt IMRPhenomD} waveform to account for the averaging over inclination angles.

\section{Error Analysis} \label{sec:Fisher}
Under the assumption of the detector noise being stationary and Gaussian, the distribution of various signal parameters can be approximated by a multivariate Gaussian described by the Fisher information matrix. 
In the limit of large SNRs, the $1\sigma$ widths provide lower limit on the statistical uncertainties associated with the measurement of the parameters usually referred to as Cramer-Rao bound~\cite{Cramer46,Rao45}.
Fisher information matrix is the noise weighted inner product of the derivatives of the frequency-domain waveform with respect to the eight parameters that we are concerned here and evaluated at the true value of the parameters. Therefore, with the knowledge of the gravitational waveform of interest and the projected sensitivity of the detector, we can predict the measurement uncertainties of the parameters. There have been criticisms of the use of Fisher matrix for such projections, especially on signals that may have SNR ${\cal O}(10)$ which is the case for LIGO and Virgo detectors \cite{vallisneri2008use}. However, if the problem in hand is to assess at the order of magnitude level the statistical uncertainties in the measurement, Fisher matrix still provides a useful method to obtain them. More rigorous methods that numerically sample the likelihood functions may be used to quantify this more precisely as a future work. For instance, a recent work \cite{Dupletsa:2024gfl} carried out such a comparison in the context of XG detectors and argued that with an appropriate choice of priors, Fisher matrix based method can be employed for assessing the performance of XG detector configurations.

For different representative binary configurations, Fisher matrix can be computed for a given detector PSD which, in our case, is a $8\times8$ symmetric matrix, by construction. Inverse of the Fisher matrix is called variance-covariance matrix. Square root of the diagonal entries of this matrix gives $1\sigma$ error bar that is of interest to us.
More precisely, the Fisher matrix is defined as
\begin{equation}
    \Gamma_{ab}=2\int_{f_{\rm low}}^{f_{\rm up}}\,\frac{{\tilde h}_{,a}{\tilde h}^{*}_{,b}+{\tilde h}_{,b}{\tilde h}^{*}_{,a}}{S_n(f)} df
\end{equation}
where commas denote partial differentiation of the waveform with respect to various parameters $\theta^a$ and asterisk denote complex conjugation. The tilde denotes the Fourier transform of the time domain signal $h(t)$ and $S_n(f)$ is the noise PSD of the detector of interest. 

In this work, we study three representative detector configurations, AdvLIGO as the representative of the second-generation GW detector~\footnote{As the designed sensitivity of Virgo is similar to that of LIGO, we use LIGO as a proxy for Virgo and any other detectors that have similar sensitivity.}, Cosmic Explorer and LISA as representatives of the ground-based and space-based next-generation detectors, respectively. We use the designed noise PSD of Advanced LIGO, given in  Eq.(4.7) of \cite{Ajith:2011ec}, the CE PSD given in \cite{Kastha:2018bcr}, the ET PSD in \cite{Hild:2010id} and the LISA noise PSD discussed in \cite{Babak:2017tow,Datta:2023muk}. The LISA noise PSD has two distinct contributions, one from the instrument noise and another from the galactic confusion noise. The instrumental noise PSD given in \cite{Babak:2017tow} is divided by a factor of 2 to account for summation over two independent frequency channels. On the other hand, the unresolved galactic binaries contribute to a background confusion noise in the low frequency regime, $f \lesssim 1$mHz, which is modelled through an analytical expression given in \cite{Mangiagli:2020rwz} for a four year observation period of LISA. 

Signal to noise ratio, or simply SNR, quantifies the strength of the signal in a detector data. The SNR denoted by $\rho$, is defined using the Fourier transform of the signal $\Tilde{h}(f)$, as
\begin{eqnarray}
    \rho^2 = 4 \int_{f_{\rm low}}^{f_{\rm up}} \frac{|\tilde{h}(f)|^2}{S_n (f)} df \,.
\end{eqnarray}

The lower cut-off frequency, for ground-based detectors, in the Fisher analysis depends on the detector, $f_{\rm low} = 10$ Hz for AdvLIGO, $5$ Hz for CE and $1$ Hz for ET. The upper cut-off frequency, on the other hand, is formally infinity. However, following Ref.~\cite{Datta:2020vcj}, we set
$f_{\rm up} = f_{\rm IMR}$ where $f_{\rm IMR}$ corresponds to the frequency at which the characteristic amplitude $2 \sqrt{f} |\tilde{h}(f)|$ of the GW signal is lower than that of the detector noise amplitude spectral density by 10\% at maximum. For binaries of mass $10-110 M_{\odot}$ at a luminosity distance of 500 Mpc, the SNR for different mass ratios vary from $\sim 10-50$ for AdvLIGO and for CE with masses $\sim 10-600$ $M_{\odot}$, the SNR varies from $\sim 10^2-10^4$. For AdvLIGO, certain mass choices, like total mass of $10$ $M_{\odot}$ with any mass ratio, has SNR $<10$ which are excluded from our analysis.

Since the sensitivity of LISA will allow the observation of GW signals from supermassive black holes, we select the mass range of the binary to be $10^4 -10^7$ $M_{\odot}$ while keeping the mass ratios same as used for the ground-based detectors. LISA is sensitive in the mHz frequency regime with lower and upper cut-off frequencies decided by the binary parameters. The lower cut-off frequency is chosen such
that the GW signal from the inspiraling binary lasts for four years prior to its merger but is not lower than the low-frequency limit of the LISA noise PSD, which is $10^{-4}$ Hz.
Hence, the lower cut-off frequency is chosen as~\cite{BBW05a,Datta:2023muk}
\begin{equation}
    f_{\rm{low}} = {\rm Max} \Big[10^{-4},\, 4.149 \times 10^{-5} \Big( \frac{{\rm M_c}}{10^6} \Big)^{-5/8} {\rm T^{-3/8}_{obs}} \Big],
\end{equation}
where ${\rm T_{obs}}$ is the duration of observation of LISA i.e. four years and ${\rm M_c}$ is the chirp mass in solar mass units. The upper cut-off frequency is chosen between $f_{\rm IMR}$ and the upper frequency limit of 0.1 Hz, whichever is smaller,
\begin{equation}
    f_{\rm up} = {\rm Min} \Big[f_{\rm IMR}, \, 0.1 \Big].
\end{equation}
The supermassive black hole binaries at 3 Gpc luminosity distance have SNR in LISA between $\sim 10^2 -10^4$ for different mass and mass ratios.

In our analysis, we also incorporate the effect of redshift in the observed masses of the compact binaries through a factor of $1+z$, where $z$ is the redshift of the source. That is, the detector-frame masses $m_{\rm det}$ are related to the source-frame masses of the binary $m_{s}$ as $m_{\rm det} = m_s (1+z)$. Throughout the paper, we mention the source-frame masses of the binaries unless specified otherwise.
For a fixed luminosity distance, we assume the flat $\Lambda$-CDM model and calculate the associated redshift $z$ by employing
\begin{equation}
\rm{d_L} (z) = \frac{(1+z) }{H_0} \int_0^z \frac{dz'}{\sqrt{\Omega_M (1+z')^3 + \Omega_\Lambda}} \,,
\label{eq:redshift}
\end{equation}
where the cosmological parameters are $\Omega_{M} = 0.3065$, $\Omega_\Lambda = 0.6935$ and $h = 0.6790$ with $\rm{H_0} =100h$ (km/s)/Mpc (\cite{Planck:2015fie}). In the Fisher matrix code, the assumed luminosity distance of the source is used to obtain the redshift of the source and hence the redshifted mass. 

\section{Results} \label{sec:result}
\subsection{Projected bounds on events like GW150914 and GW151226}

In this section, we show the projected bounds on all the deformation parameters starting from 0PN to 4.5PN with AdvLIGO sensitivity for binary black hole having parameters similar to GW150914 \cite{LIGOScientific:2016aoc} and GW151226 \cite{GW151226}, the first two detections made by LIGO during the first observing run. On the one hand, these two events represent two interesting regimes of the dynamics. GW150914 is a relatively high mass system for which we observe only a few cycles of late inspiral whereas GW151226 has several cycles of inspiral in the frequency bands of LIGO/Virgo. This also helps understand how the errors vary as a function of PN order. As the LIGO-Virgo-KAGRA (LVK) collaboration analyses usually quote 90\% credible bounds, we convert the $1\sigma$ bounds from Fisher matrix to 90\% credibility. The parameters of the binaries are taken from the median values of the LVK posteriors, including those of the luminosity distances (see Table III of \cite{LIGOScientific:2018mvr}). 

The projected bound on the 12 deformation parameters for AdvLIGO-like sensitivity are shown in Table \ref{tab:GW151226} for these two binaries. The variation of the bounds across PN orders does not show any monotonic trends. Parameters at higher PN orders are not necessarily more poorly constrained than some of the lower PN order parameters. This is due to the well-known oscillatory convergence of the PN series and has been observed in various data analysis contexts (see for example Table 1 of Ref.~\cite{Arun:2004hn}). 
The trends till 3.5PN can be compared against the trends reported by LVK in Fig. 4 of \cite{GWTC1TGR:2019fpa} from the analysis of the two  above mentioned GW events. We find these two trends to match exactly. We cannot compare the bounds themselves here as, apart from using the Fisher matrix for the projections, the noise PSDs we use are that of the designed sensitivity of AdvLIGO, whereas \cite{GWTC1TGR:2019fpa} uses the sensitivity of LIGO and Virgo during the first observing run when these two events were detected.  Precisely due to this reason, our bounds are better than those in \cite{GWTC1TGR:2019fpa}. Next, looking at the bounds on the new parameters that appear at 4PN and 4.5PN orders, we find with the exception of $\delta \hat{\phi}_{8\ell}$, the other three parameters are likely to yield poorer constraints than all the parameters till 3.5PN.

\begin{table}[h]
 \begin{tabular}{|c|c|c|}
      \hline
     $\delta \hat{\phi}_k$& GW150914-like & GW151226-like \\ [0.1 cm]
    \hline 
   $\delta \hat{\phi}_0$ &0.05& 0.18 \\[0.1 cm]
   $\delta \hat{\phi}_2$ &0.11& 0.14 \\[0.1 cm]
   $\delta \hat{\phi}_3$ &0.06& 0.13 \\[0.1 cm]
   $\delta \hat{\phi}_4$ &0.41& 1.21 \\[0.1 cm]
   $\delta \hat{\phi}_{5\ell}$ &0.13& 0.35 \\[0.1 cm]
   $\delta \hat{\phi}_6$ & 0.25& 0.92 \\[0.1 cm]
   $\delta \hat{\phi}_{6\ell}$ &0.98& 2.34\\[0.1 cm]
   $\delta \hat{\phi}_7$ & 0.50& 1.46\\[0.1 cm]
   $\delta \hat{\phi}_{8\ell}$ &0.13 & 0.31 \\[0.1 cm]
   $\delta \hat{\phi}_{8\ell^2}$ &1.16& 3.63 \\[0.1cm]
   $\delta \hat{\phi}_9$ & 1.87& 3.12 \\[0.1cm]
   $\delta \hat{\phi}_{9\ell}$ &5.31& 8.17 \\[0.1cm]
   \hline
\end{tabular}
\caption{Projected Fisher 90\% bound obtained for GW150914-like (at 440 Mpc) and GW151226-like (at 450 Mpc)  binary masses and spins with AdvLIGO sensitivity having SNR $39.8$ and $16.2$ respectively. We use our modified waveform which includes all deformation parameters till $4.5$PN.}
\label{tab:GW151226}
\end{table}

\begin{figure*}
    \includegraphics[width=1\textwidth]{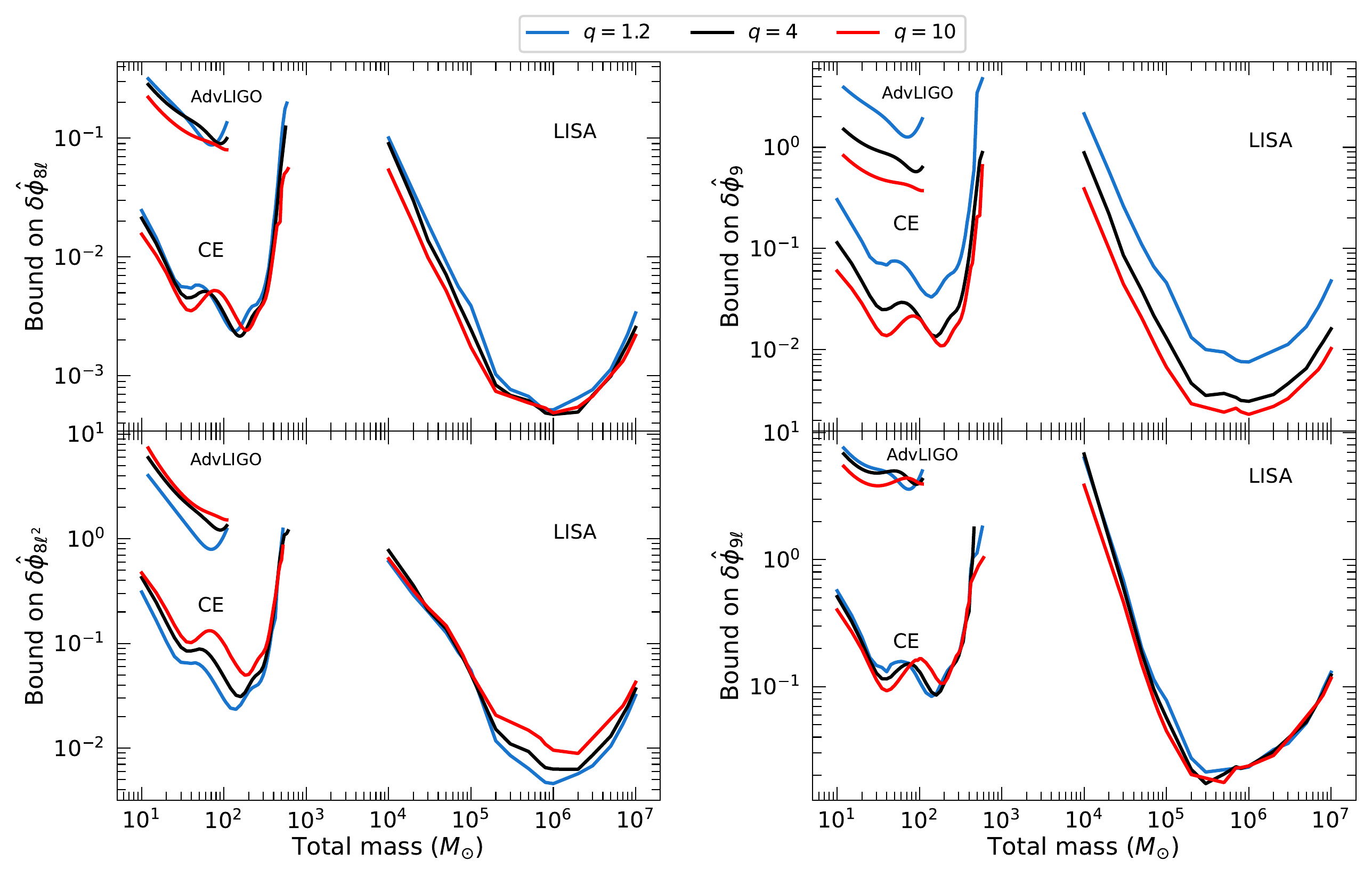}
    \caption{Projected $1\sigma$ bounds on the deformation parameters corresponding to the 4PN and 4.5PN phasing terms for AdvLIGO, CE and LISA sensitivities. Different total mass binaries are considered with mass ratios $q=1.2,4,10$ and aligned spin $\chi_{1,2} =(0.2,0.1)$. The sources are at a luminosity distance of 500 Mpc ($z=0.1049$) for AdvLIGO/CE and at 3 Gpc ($z=0.512$) for LISA.}
    \label{fig:bounds}
\end{figure*}

\subsection{Variation of the bounds as a function of binary parameters}
We now compute the projected bounds on $\{\delta \hat{\phi}_{8\ell},\, \delta \hat{\phi}_{8\ell^2},\, \delta \hat{\phi}_{9}, \,\delta \hat{\phi}_{9\ell}\}$, and their variation as a function of total mass and mass ratios of the binaries for current generation detectors (represented by AdvLIGO), next-generation ground-based detector (represented by CE) and space-based LISA detector and analyse masses from a few solar masses to $10^7M_{\odot}$. Since our work focuses on constraining these deformation parameters associated with the non-spinning part of the inspiral phase, we keep the magnitudes of the aligned binary spins fixed in our analysis. The spins are chosen to be $(0.2,0.1)$, which is consistent with the fact that the observed BBH population has relatively smaller component spins~ \cite{GWTC3:2021sio}. The variation in spin magnitude is not expected to alter the trends shown by the bounds significantly. A detailed quantification of this will be addressed in a future work.  
\subsubsection{Advanced LIGO and Cosmic Explorer results}
We focus on the bounds of the deformation parameters from ground based detectors in this section. We consider the binaries to be at $500$ Mpc ($z=0.1049$) having spin magnitudes $(0.2,0.1)$ aligned with the orbital angular momentum. The total mass is varied from $10 - 110$ $M_{\odot}$ for AdvLIGO whereas from 10 - 600 $M_{\odot}$ for CE. There is very little inspiral in the bands of the respective detectors beyond these mass ranges and hence binaries with masses after this maximum mass are unsuitable for our tests. For a particular total mass, we study systems with different mass ratios $q=1.2,4,10$. Figure~\ref{fig:bounds} shows the $1\sigma$ bound on the deformation parameters as a function of the total mass of the binary for different $q$ for both AdvLIGO and CE.

Initially, we observe a gradual improvement in the bound of the parameters with increasing total mass, for all detectors, that can be attributed to increasing SNR for high-mass systems. But as the total mass increases, the signal has lesser number of cycles in band as the merger frequency is inversely proportional to the mass. This leads to a degradation of the bounds after some total mass depending on the detector PSD. Of the four deformation parameters, $\delta \hat{\phi}_{8\ell}$ has the best bound with a precision of $\sim 10^{-1} (10^{-3})$ for AdvLIGO (CE). Further, we find CE can constrain $\delta \hat{\phi}_{9}$ by $\sim 10^{-2}$ and the remaining deformation parameters by $\sim 10^{-1}$. Typical numbers for representative systems (GW150914-like and GW151226-like) are also shown in Fig.~\ref{fig:summary}. We observe, once again, in Fig.\ref{fig:summary}, the higher mass binary corresponding to GW150914 gives better bound on the parameters than GW151226, with $\delta \hat{\phi}_{8\ell}$ being best measured.
\begin{figure*}
    \includegraphics[width=1\textwidth]{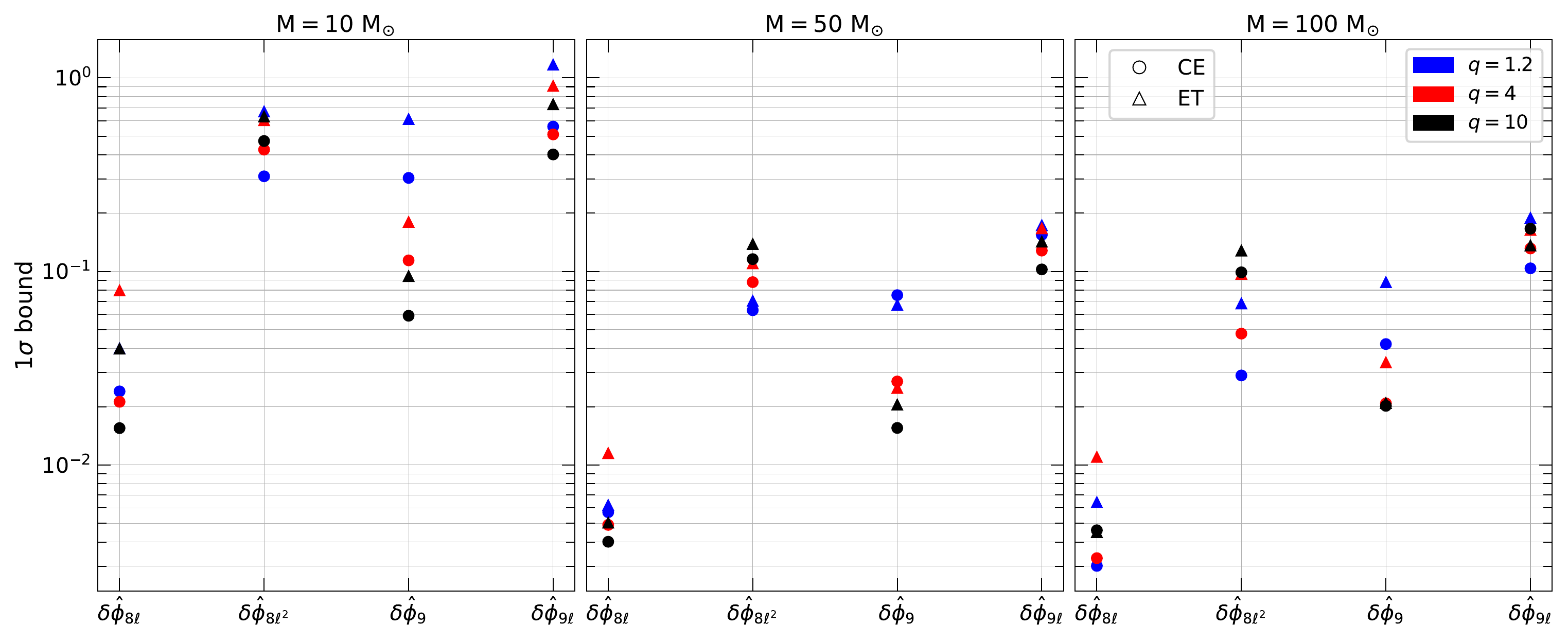}
    \caption{Projected $1\sigma$ bounds on the deformation parameters corresponding to the 4PN and 4.5PN phasing terms for CE and ET sensitivities. Binaries of total mass 10, 50 and 100 $M_{\odot}$ are considered as representatives with mass ratios $q=1.2,4,10$ and aligned spin $\chi_{1,2} =(0.2,0.1)$. The sources are at a luminosity distance of 500 Mpc ($z=0.1049$). Bounds from CE and ET are comparable for higher mass binaries while CE performs better for low mass cases.}
    \label{fig:Compare}
\end{figure*}
Figure~\ref{fig:bounds} shows that the bound on $\delta \hat{\phi}_{8\ell}, \delta \hat{\phi}_{9}$ and $\delta \hat{\phi}_{9\ell}$ improves for more asymmetric binaries or systems which has larger mass ratios. This is because, the PN coefficients corresponding to these three terms are dominated by the non-quadrupolar modes~\cite{Blanchet:2023,PhysRevD.108.064041} which are strongly excited for the asymmetric systems hence leading to better bound for asymmetric binaries. On the other hand, bounds on $\delta \hat{\phi}_{8\ell^2}$ are the best when the binary is more symmetric. Upon examining the multipoles that contribute to the 4PN log-square term, it is evident that quadrupole moment occurring at different PN orders dominantly contribute to this order, hence giving better bounds for symmetric binaries. 
   
Finally, it is seen from the plots that with the exception of $\delta \hat{\phi}_{9\ell}$, parameters at $4$PN and $4.5$PN can yield bounds $\leq {\cal O}(1)$, even with AdvLIGO sensitivity for an appropriate ranges of mass and mass ratios, presenting a unique opportunity to test the validity of GR at such high PN orders even with LIGO. The magnitude of the bound improves significantly for CE due to its enhanced sensitivity.

Since the bounds in Fig.\ref{fig:bounds} are computed only with CE sensitivity for the case of 3G detectors, we compute the bounds for ET as well for certain representative masses and compare them with CE. In Fig.\ref{fig:Compare}, we show bounds as obtained from both CE and ET for 10, 50 and 100$M_{\odot}$ and varying mass ratios. We observe that with increasing total mass, the projected bounds from the two detectors become comparable due to the lower number of inspiral cycles. On the other hand, for $10 M_{\odot}$, CE performs slightly better since the sensitivity of CE is better in the mid-frequency range than ET where majority of the inspiral cycles fall. Overall, the conclusions obtained from Fig.\ref{fig:bounds} still hold when ET is included in the analysis.

\subsubsection{Results for LISA}
In the previous section, we observed that the four new deformation parameters at 4 and 4.5PN can be bounded reasonably with AdvLIGO and CE sensitivity. The total mass of the binaries that yield the best bounds are $\sim 100 M_{\odot}$. On the other hand, the space-based detector LISA will be observing the merger of supermassive black hole binaries with masses of $\sim 10^4-10^7 M_{\odot}$. In this section, we will estimate the projected bounds on $\{\delta \hat{\phi}_{8\ell},\, \delta \hat{\phi}_{8\ell^2},\, \delta \hat{\phi}_{9}, \,\delta \hat{\phi}_{9\ell}\}$ from the GW signals of supermassive black hole binary mergers that will be detected by LISA. The total mass of the binaries is varied from $10^4-10^7$ $M_{\odot}$ and the mass ratios are $q=1.2,4,10$. We consider the binaries at a prototypical luminosity distance of 3 Gpc ($z=0.512$), with spin magnitudes $(0.2,0.1)$ aligned with the orbital angular momentum. Fig.\ref{fig:bounds} shows our LISA results.

The dependence of the bounds on the mass ratios are qualitatively the same as observed for AdvLIGO/CE. Similar to the case of AdvLIGO/CE, the best bounds on the parameter $\delta \hat{\phi}_{8\ell^2}$ are for the more symmetric binaries with $q=1.2$. The remaining three parameters are better constrained for more asymmetric systems. 
The parameter $\delta \hat{\phi}_{8\ell}$, once again, has the best bound of $\sim 10^{-3}$ while \{$\delta \hat{\phi}_{8\ell^2},\, \delta \hat{\phi}_{9}$\} have bounds of $\sim 10^{-2}$ and finally, $\delta \hat{\phi}_{9\ell}$ has the worst bound of $\sim 10^{-1}$. Hence, we find the four deformation parameters can be constrained to very good precision with GWs from supermassive
binary black holes, as observed by LISA. These bounds clearly outsmart the bounds from CE, thanks to the longer duration of the signals in the LISA band.

\section{Conclusion} \label{sec:conclusion}
The parametrized tests of GR using the inspiral dynamics are currently performed using the expansion of the inspiral phase up to 3.5PN. The recent analytical computation of terms occurring at 4 and 4.5PN of inspiral phase for quasi-circular, non-spinning binaries allows us to extend these tests to 4.5PN. The four new PN coefficients that occur at 4 and 4.5PN permits tests of novel physical effects such as the tail-of-memory, spin-quadrupole tails and quartic tails. In this work, we compute the projected $1\sigma$ bounds on the four new deformation coefficients,$\{\delta \hat{\phi}_{8\ell},\, \delta \hat{\phi}_{8\ell^2},\, \delta \hat{\phi}_{9}, \,\delta \hat{\phi}_{9\ell}\}$ which are introduced in the logarithmic, square-logarithmic and non-logarithmic terms appearing at 4 and 4.5PN. We employ Fisher analysis with modified {\tt IMRPhenomD} waveform for estimating the bounds. For different binary configurations and detectors (AdvLIGO, CE and LISA), the bounds are shown in Fig.\ref{fig:bounds} and the main results are summarized as follows.

The parameter corresponding to 4PN log-term, $\delta \hat{\phi}_{8\ell}$ has the best bound of $\sim 10^{-3}$ from LISA, $\sim 10^{-2}$ from CE and $\sim 10^{-1}$ from AdvLIGO. For the remaining three deformation parameters, the bounds are $\mathcal{O}(10^{-2}-1)$ for CE and $\leq\mathcal{O}(10)$ for AdvLIGO. 
The best constraint for all the parameters are obtained from supermassive binary black holes observed in LISA due to the longer duration of the inspiral signal observed in its band. 

The network of 3G detectors will be observing orders of magnitude more number of sources compared to the current generation detectors. This allows one to combine the bounds from these events. One can do this either by multiplying the respective likelihoods (assuming the deformation parameter take the same value across all events)\cite{DelPozzo:2011pg, LIGOScientific:2019fpa} or hierarchically combine the posteriors allowing the deformation parameter to be different across events\cite{GWTC3:2021sio,LIGOScientific:2020tif,Isi:2019asy}. For Gaussian noise, when multiplying the likelihoods, the statistical error decreases as $\sim 1/\sqrt{N}$ where $N$ is the number of events detected, the bounds are expected to improve when combining results from multiple events. Assuming a network of 3G detectors, consisting of two CE with arm lengths of 40km and 20km and one ET, $N \sim 6 \times 10^4$ BBH events with SNR greater than 30 are expected to be observed per year \cite{Gupta:2023lga}. With these estimates, the bounds on the 4PN log term, say, will improve from $\mathcal{O}(10^{-2})$ to $\mathcal{O}(10^{-5})$ for 3G. Similar estimations for LISA is difficult due to the uncertainties related to the detection rates of SMBBH mergers by LISA.

Finally, we conclude that apart from $\delta \hat{\phi}_{9\ell}$, deviation from GR at 4PN and 4.5PN can be constrained with $\leq \mathcal{O}(1)$ precision even with AdvLIGO sensitivity,
presenting a unique possibility to utilize the rich characters of the inspiral phase in the high freqeuncy regime to study GR violation.

All these projections are based on Fisher matrix formalism and which are valid when the SNRs are sufficiently high. In order to assess the error bars on our projections, we compared the bounds from the parametrized tests performed for GW150914 and GW151226 with  what our approach would have predicted for the same. The results are tabulated in Table~\ref{tab:comparison} and details of the comparison are given in Sec.~\ref{sec:appendixb} of the Appendix. For inspiral-dominated GW151226, our projections are found to underestimate the true bounds up to a factor of 2. This underestimation can be up to a factor of 4 (or even 8 for 2.5PN logarithmic parameter) in the case of more massive GW150914. Therefore, a more detailed Bayesian analysis with waveforms having precession and higher modes will be the next step to support the bounds from Fisher analysis and will be pursued through future projects.

\section*{Acknowledgements}
The authors thank Sebastian Khan for sharing his Mathematica code
of the IMRPhenomD waveform model with us. We thank N. V. Krishnendu for useful comments on the manuscript. The authors also thank Pankaj Saini and Parthapratim Mahapatra for useful discussions. K.G.A. acknowledges Swarnajayanti Fellowship Grant No.~DST/SJF/PSA-01/2017-18 and Core Research Grant No.~CRG/2021/004565  of the SERB. K.G.A and P.D.R. acknowledges the support from Infosys foundation. S.D. acknowledges support from UVA Arts and Sciences Rising Scholars Fellowship. This material is based upon work supported by NSF's LIGO Laboratory which is a major facility fully funded by the National Science Foundation. This manuscript has the LIGO preprint number {\tt P2400185.}

\appendix

\begin{table}[h]
  \begin{tabular}{|c|c|c|c|c|}
      \hline
     Event & PN & LVK & Fisher $90\%$ bound & LVK/Fisher\\ [0.1 cm]
    \hline 
   GW150914 &  0 & 0.2 & 0.07 & 2.57\\[0.1 cm]
   SNR$=16.311$ &    1 & 0.6 & 0.21 & 2.88\\ [0.1 cm]
   LVK SNR$=25.3$ &1.5 & 0.4 & 0.12 & 3.39\\ [0.1 cm]
       &2 & 3 & 0.61 & 4.94\\ [0.1 cm]
   &2.5 l & 1.5 & 0.18 & 8.23\\ [0.1 cm]
   & 3 & 2 & 1.07 & 1.85\\[0.1 cm]
   &   3 l & 10.5 & 4.13 & 2.54\\[0.1 cm]
   &   3.5 & 5.5 & 3.05 & 1.79\\ [0.1 cm]
  \hline
   GW151226 &   0 & 0.2 & 0.13 & 1.55\\[0.1 cm]
   SNR$=6.38$&     1 & 0.3 & 0.23 & 1.33\\[0.1 cm]
   LVK SNR$=12.4$     & 1.5 & 0.2 & 0.14 & 1.39\\[0.1 cm]
      & 2 & 1.8 & 1.09 & 1.65\\[0.1 cm]
   & 2.5l & 0.6 & 0.37 & 1.61\\[0.1 cm]
   &      3 & 1.5 & 1.19 & 1.61\\[0.1 cm]
   &     3l & 7 & 4.16 & 1.68\\[0.1 cm]
   &     3.5 & 4 & 2.28 & 1.74\\[0.1 cm]
    \hline
 \end{tabular}
\caption{Comparison of 90\% bounds on deformation parameters using O1 noise PSD and {\tt IMRPhenomD} waveform till $3.5$PN for GW150914 and GW151226. We have taken the values of binary mass, mass ratio, spin and luminosity distance, for the events, as quoted in Table III of \cite{LIGOScientific:2018mvr}. The Fisher matrix bound is normalised with the network SNR corresponding to these events.}
\label{tab:comparison}
\end{table}


\section{Comparison of Fisher-based bounds with existing LVK results} \label{sec:appendixb}
The Fisher matrix projections are expected to be reliable only for high SNR systems. In this section, we will compare the bounds on the deviation parameters, till $3.5$ PN, computed through the Fisher matrix with those obtained from Bayesian analysis to estimate the accuracy of our predicted bounds for the new deformation parameters. More specifically, we compare the Fisher-based 90$\%$ bound with the values obtained by LVK in the events catalogued in GWTC-1\cite{LIGOScientific:2018mvr}. We take $f_{\rm low} =20$ Hz and the upper limit of frequency being $f_{\rm IMR}$ (see Sec.\ref{sec:Fisher}). The $1\sigma$ bounds obtained from Fisher analysis are converted to those at 90\% credibility .
We use the O1 noise PSD and normalize the Fisher matrix bound with respect to the network SNR corresponding to a particular event. We approximate the $\chi_{1,2}$ values by the median $\chi_{\rm eff}$ value quoted in \cite{LIGOScientific:2018mvr}. Likewise, we consider the corresponding median values for the mass and luminosity distance of the binaries. The last column in Table \ref{tab:comparison} shows the LVK to Fisher bound ratio, which, if equal to 1, denotes the exact match between the bounds from the two methods. We find the Fisher-based bounds comparable with the Bayesian bounds, at least for GW151226 which is inspiral-dominated. Even for heavy mass binaries like GW150914, apart from 2.5PN log-term, the constraint on the other PN order deformation parameters differ from the LVK bound by a factor of $\sim 1$ to 4.

\bibliographystyle{apsrev}
\bibliography{refs}
\end{document}